\documentclass[3p,times,procedia]{elsarticle}
\usepackage{nupha_ecrc}

\volume{00}

\firstpage{1}

\journalname{Nuclear Physics A}

\runauth{}

\jid{nupha}

\jnltitlelogo{Nuclear Physics A}

\usepackage{amssymb}

\usepackage{xcolor}
\usepackage{hyperref}
\usepackage[figuresright]{rotating}

\begin{document}

\begin{frontmatter}



\dochead{XXVIIIth International Conference on Ultrarelativistic Nucleus-Nucleus Collisions\\ (Quark Matter 2019)}

\title{New developments in lattice QCD on\\ equilibrium physics and phase diagram}


\author{Heng-Tong Ding}
\address{Key Laboratory of Quark \& Lepton Physics (MOE) and Institute of Particle Physics, \\Central China Normal University, Wuhan 430079, China}

\begin{abstract}
	I review recent new lattice QCD results on a few selected topics which are relevant to the heavy ion physics community.  Special emphasis is put on the QCD phase diagram towards the chiral limit and at nonzero baryon density as well as the fate of quarkonia and heavy quark drag coefficients.
\end{abstract}


\end{frontmatter}



\section{Introduction}

Lattice QCD is a technique based on first principles in which strong interaction quantities are calculated by large-scale numerical Monte-Carlo simulations of the Euclidean space path-integral in a finite-volume discretized space-time, where the effects of the finite volume and the discretization can be systematically removed in the thermodynamic limit and the continuum limit.
Several milestones have been achieved in the past in the lattice QCD simulations, e.g. QCD transition at physical point at vanishing baryon density is not a true phase transition but a rapid crossover~\cite{Aoki:2006we}, equation of state of $N_f=2+1$ QCD is obtained in the continuum limit~\cite{Bazavov:2014pvz,Borsanyi:2013bia} {\it etc}. In this proceedings I will focus on the QCD phase structure at physical point and towards the limit of massless pions,  and their connections to the location of QCD critical end point at nonzero baryon density $\mu_B$, and the higher order baryon number fluctuations at small values of $\mu_B$ as well as the fate of quarkonia states and heavy quark drag coefficients.

\section{QCD transition towards the chiral limit}
One of the main goals of Lattice QCD as well as heavy ion community is to map out the QCD phase diagram, 
practically to find the elusive QCD critical end point (CEP) or a critical region~\cite{Ding:2015ona,Luo:2017faz}. 
This critical end point is supposed to be a 2nd order phase transition belonging to a Z(2) universality class. 
On the lattice QCD side the study is hampered by the sign problem, and the most commonly-used practical approaches 
to circumvent it are the Taylor expansion method~\cite{Allton:2002zi,Gavai:2003mf} and 
simulations at imaginary chemical potential~\cite{DElia:2002tig}. On the experimental side 
the observed non-monotonic behavior of kurtosis ratios of proton number fluctuations, 
i.e. 4th to 2nd order cumulant ratios, as well as the sign change of the 6th order cumulant ratios at beam energies of 200 and 54.4 GeV have triggered tremendous interests~\cite{Luo:2017faz,STARcumulants,Adam:2020unf}.
Theoretical understanding, and in particular, a baseline on the thermal equilibrium physics provided by Lattice QCD simulations is essential.

\begin{figure}[htp!]
	\centering
	\includegraphics[width=0.4\textwidth]{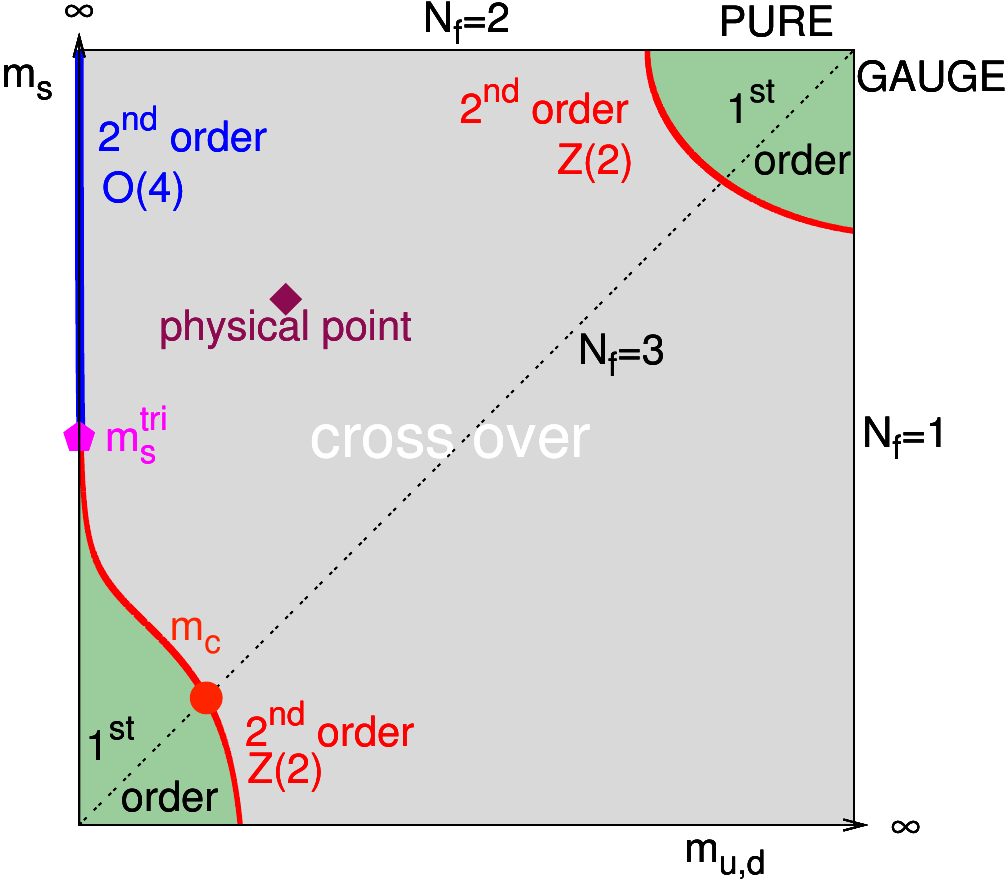}
	~~~\includegraphics[width=0.4\textwidth]{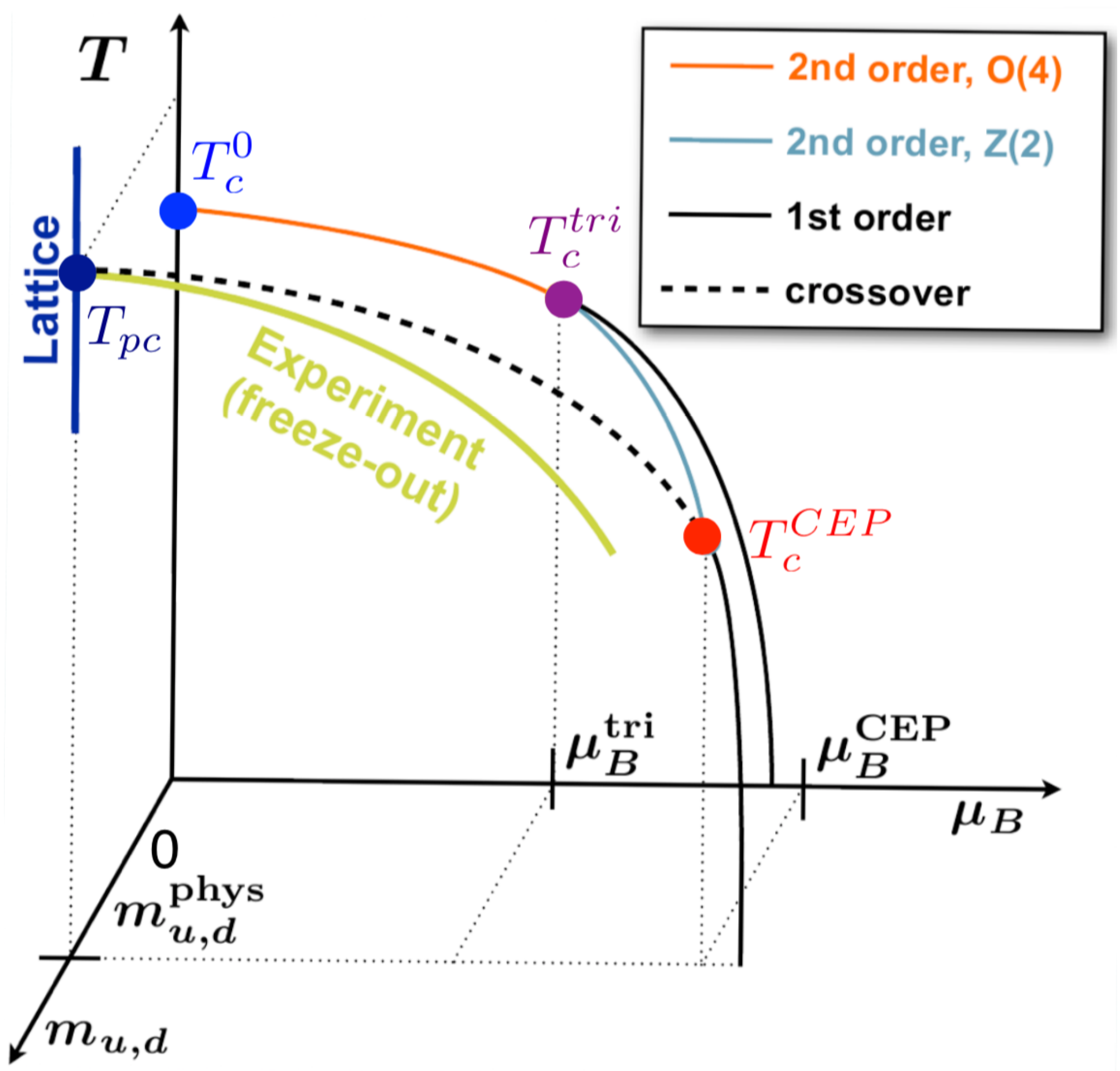}
	\caption{Left: QCD phase structure in the quark mass  plane~\cite{Ding:2015ona}. Right: QCD phase structure in the 3-D plane of temperature ($T$), quark mass ($m_{u,d}$) and baryon chemical potential ($\mu_B$).
	}
	\label{fig:QCDPhaseDiagram}
\end{figure}

While there is no criticality at zero baryon chemical potential in the real world,  
there exists several criticalities as indicated in the so-called Columbia plot (left plot of Fig.~\ref{fig:QCDPhaseDiagram}) in which the x axis is u,d quark mass while the y axis is the strange quark mass. In chiral limit of 3-flavor QCD, there is a first order chiral phase transition region which is separated from the crossover region by a 2nd order Z(2) transition line. The location of the critical line denoted by critical quark mass $m_c$, in the lower-left corner, has been studied recently on the lattice~\cite{Endrodi:2007gc,Bazavov:2017xul,Kuramashi:2020meg,Philipsen:2019rjq}. The corresponding critical pion mass becomes smaller as one approaches continuum QCD, i.e. by decreasing lattice spacing or using improved actions. The lowest critical pion mass is about 55 MeV~\cite{Bazavov:2017xul}. Thus most likely the influence from the criticality of this 1st order chiral phase transition region is not much relevant to the thermodynamics at the physical point. While on the other hand the order of phase transition in the chiral limit of a two flavor theory depends on whether the axial U(1) symmetry is effectively restored or not. Recent lattice QCD computations suggest that the transition of $N_f=$2+1 QCD in the chiral limit of light quark masses tends to be  a 2nd order transition belonging to O(4) universality
class~\cite{Fukaya:2017wfq,Suzuki:2020rla,Ding:2019fzc}.  Thus the 2nd order O(4) chiral phase transition is most relevant to the CEP at nonzero baryon chemical potential.

As sketched in the right plot of Fig.~\ref{fig:QCDPhaseDiagram} the 2nd order O(4) phase transition line in the chiral limit is supposed to end at a tri-critical point, which is connected to the CEP we are looking for, via a 2nd order Z(2) transition line.
It has been suggested from model studies that the difference of transition temperatures at tri-critical point and CEP is proportional to a power of quark mass, $T_c^{tri}-T_c^{CEP}(m_q)\propto m_q^{2/5}$~\cite{Hatta:2002sj,Halasz:1998qr,Buballa:2018hux}. This is to say that the transition temperature at the tri-critical point, $T_c^{tri}$ is larger than that at the CEP. And from lattice QCD computations the curvature of the 2nd order O(4) phase transition (orange) line is negative, i.e. the chiral phase transition temperature decreases with increasing  $\mu_B$~\cite{Kaczmarek:2011zz,Hegde:2015tbn}, suggesting $T_c^0>T_c^{tri}$. Based on these two points it is indicated that 
\begin{eqnarray}
T_c^{CEP}<T_c^{tri} < T_c^0, \nonumber
\end{eqnarray}
i.e. the chiral phase transition temperature $T_c^0$ is a possible upper bound of the transition temperature at the CEP. Thus the determination of chiral phase transition temperature will be helpful to constrain the location of the CEP.

\begin{figure}[hbtp!]
	\centering
	\includegraphics[width=0.45\textwidth]{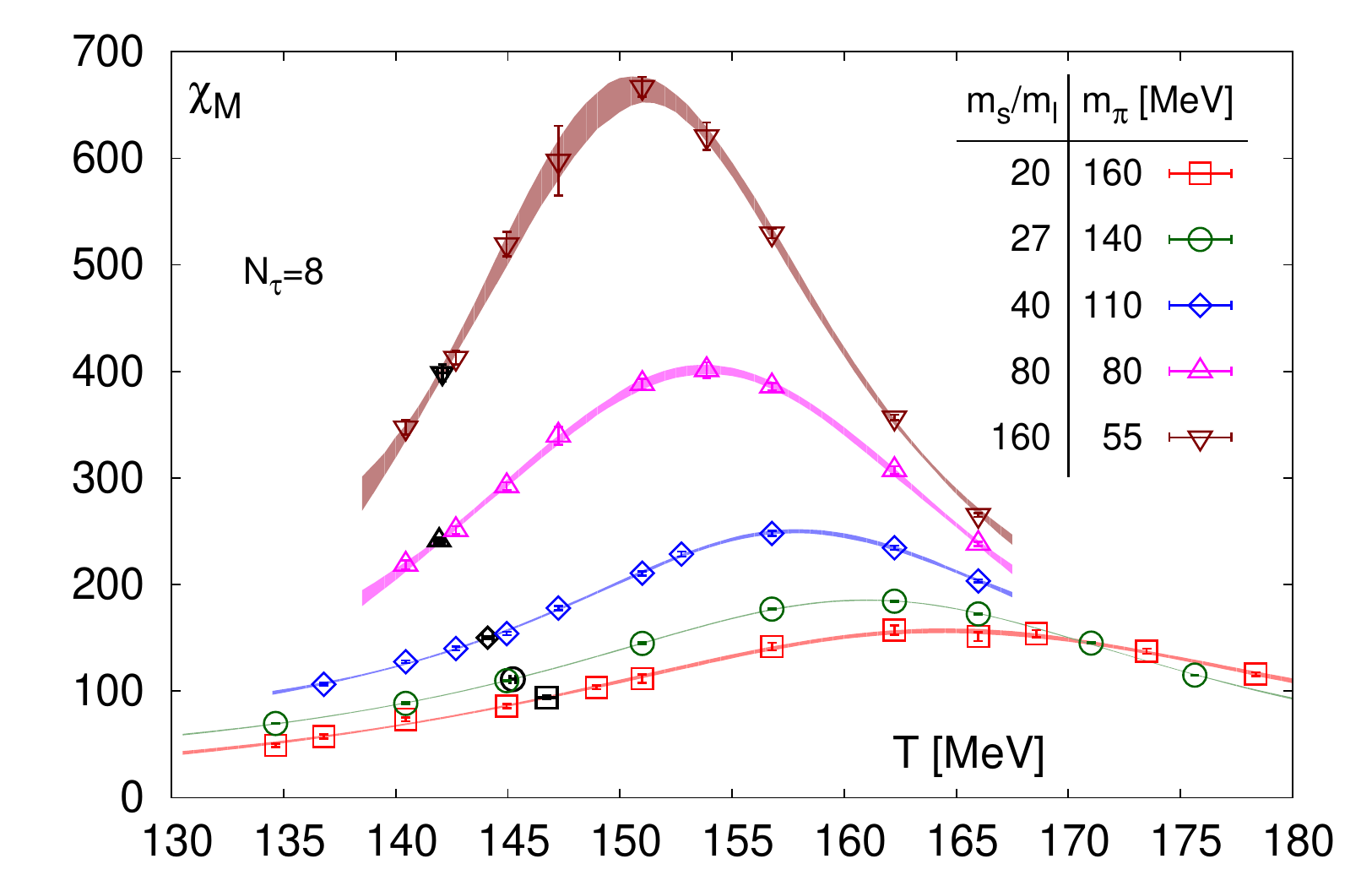}~~
	\includegraphics[width=0.38\textwidth]{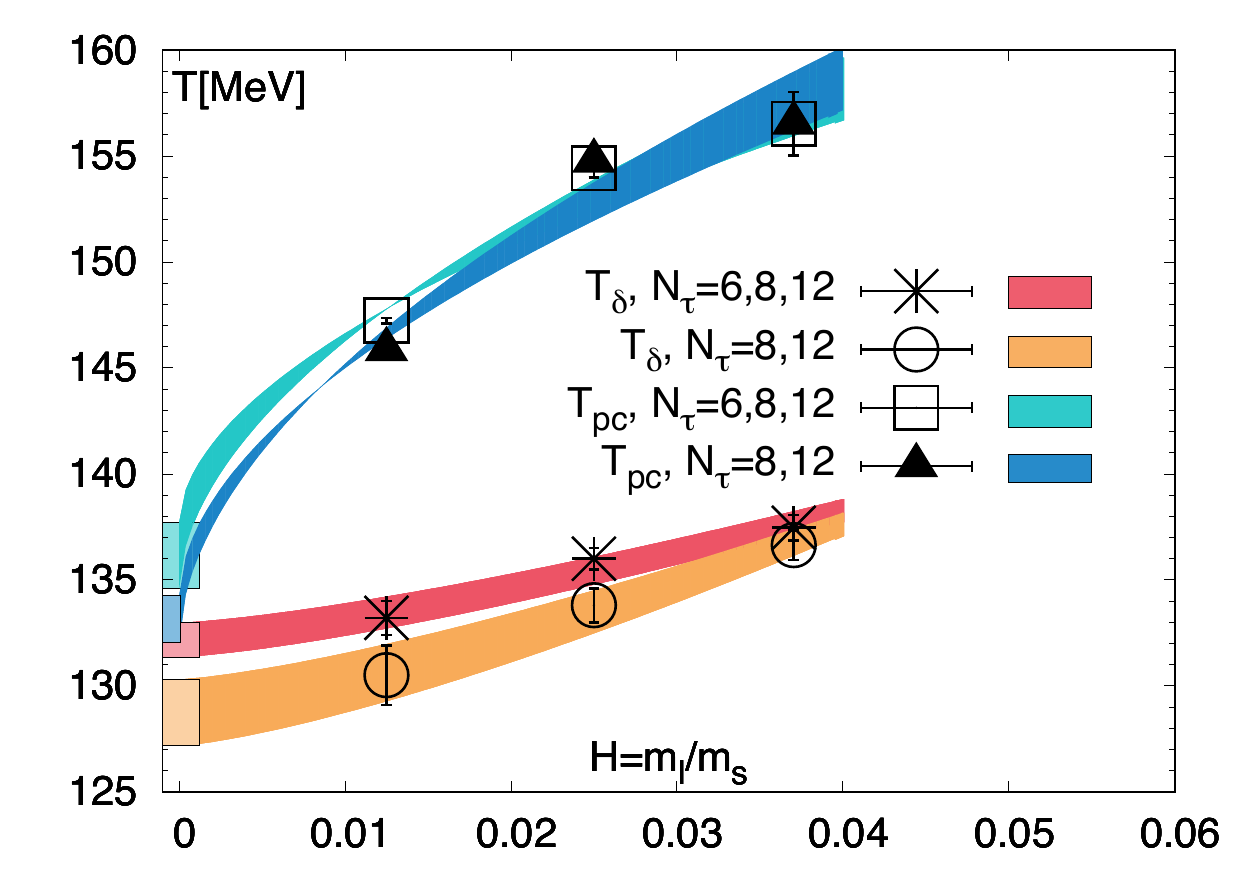}
	\caption{Determination of chiral phase transition temperature $T_c=132^{+3}_{-6}$ MeV~\cite{Ding:2019prx}. Left: Chiral susceptibility obtained from (2+1)-flavor QCD with various quark masses as a function of temperature. Right: Extrapolation of the transition temperature to the chiral limit.}
	\label{fig:chiralTc}
\end{figure}

The determination of the chiral phase transition temperature has been carried out very recently~\cite{Ding:2019prx}. As shown in the left plot of Fig.~\ref{fig:chiralTc} the maximum value of chiral susceptibility at each pion mass increases as the pion mass decreases from 160 MeV to 55 MeV. The increasing behavior of the peak height is consistent with that from the scaling relation in an O(4) universality class. Based on the O(4) scaling relations, thermodynamic limit, chiral limit as well the continuum limit have been performed~\cite{Ding:2019prx}. The extrapolation to chiral limit is demonstrated in the right plot of Fig.~\ref{fig:chiralTc}. Here the square and triangle data points give the continuum extrapolated results of chiral crossover temperature $T_{pc}(H)$ at different quark masses ratios $H$, where $T_{pc}\sim 156$ MeV at $H\simeq1/27$ stands for the  physical masses of u,d quarks and $T_{pc}(H)$ decreases with decreasing $H$.  Extrapolating to the chiral limit with the O(4) scaling relations, $T_{pc}(H)$ and an estimator ($T_\delta$) of the chiral phase transition temperature results in $T_{c}^0=132^{+3}_{-6}$ MeV~\cite{Ding:2019prx}. Since many 8th order Taylor expansion coefficients turn out to be negative in the temperature window of [135, 140] MeV~\cite{Bazavov:2020bjn,Karsch:2019mbv}\footnote{Estimates of the location of the critical end point through the radius of convergence by searching singularities in the complex plane of baryon chemical potential due to negative Taylor expansion coefficients have been put forward very recently, see e.g.~\cite{Mukherjee:2019eou,Giordano:2019gev}.}, this supports the indication of $T_c^{CEP}<T_c^{tri} < T_c^0$ and $T_c^{CEP}< 135-140$ MeV.

\section{QCD transition at the physical point and small $\mu_B$}

\begin{figure}[hbtp!]
	\begin{center}
		\includegraphics[width=0.4\textwidth]{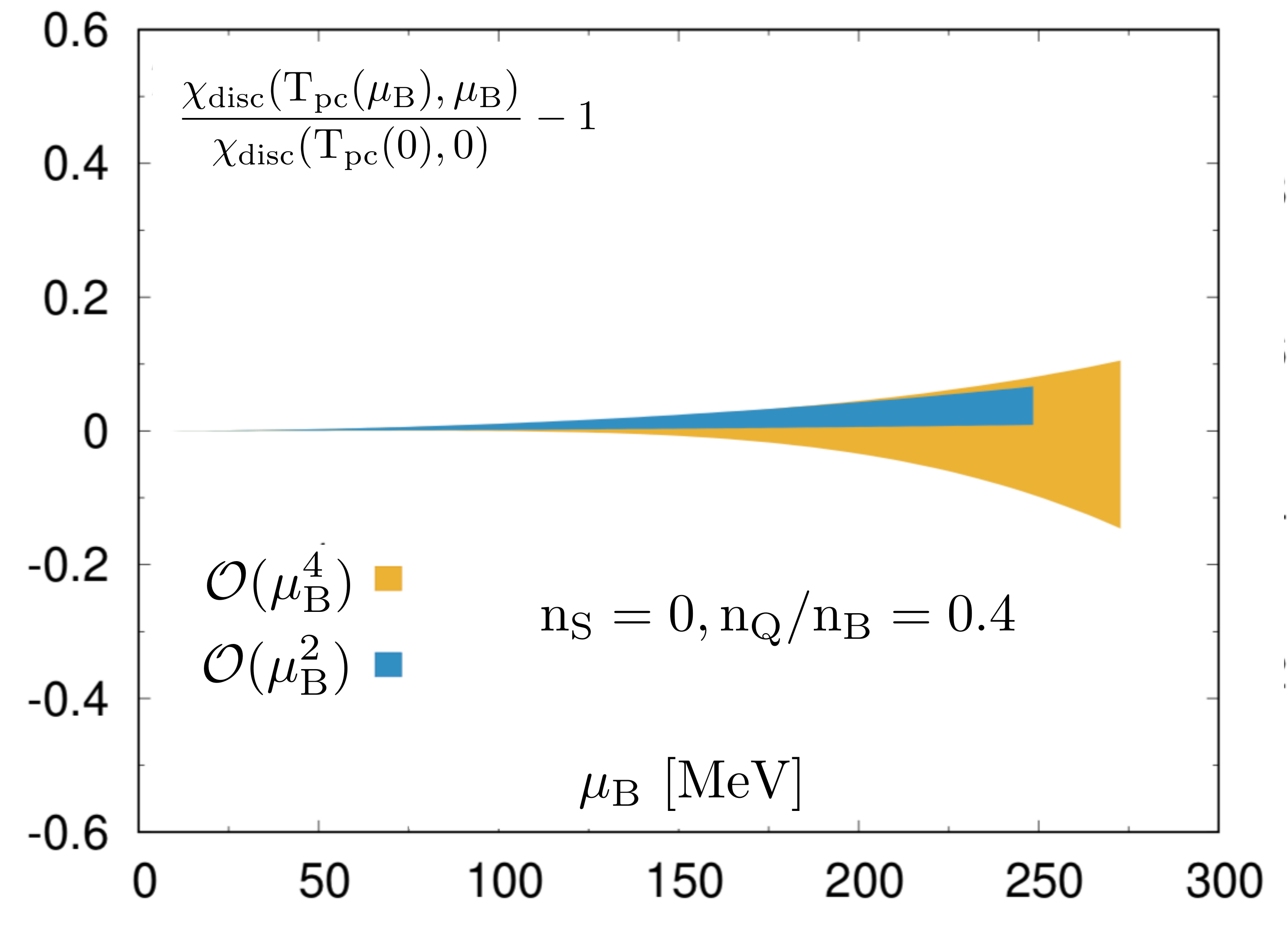}~~~~~
		\includegraphics[width=0.4\textwidth]{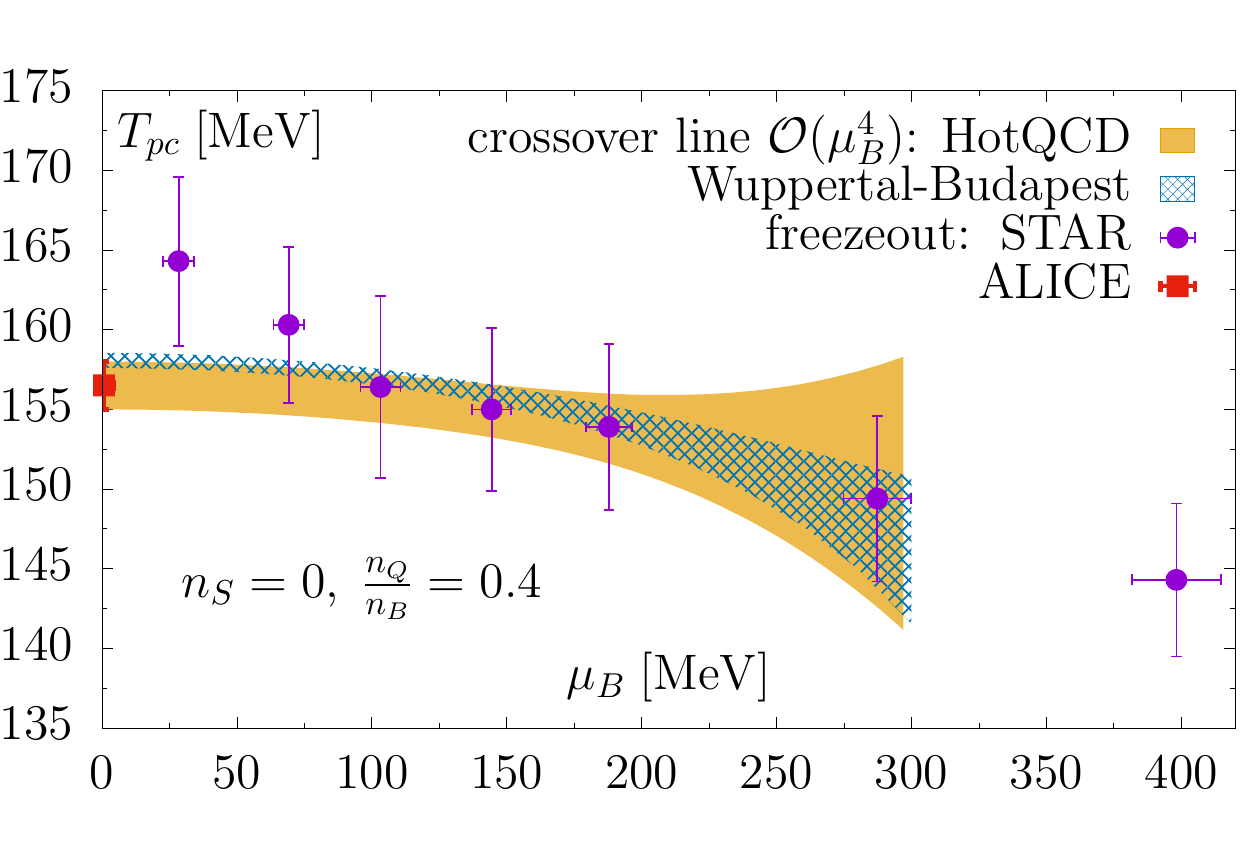}
		\caption{Left: Difference of disconnected chiral susceptibilities at nonzero and zero $\mu_B$ along the chiral crossover transition line~\cite{Bazavov:2018mes}. Right:
			Chiral crossover temperature as a function of baryon chemical potential obtained up to the order of $(\mu_B/T)^4$ from the HotQCD~\cite{Bazavov:2018mes} and Wuppertal-Budapest collaborations~\cite{Borsanyi:2020fev}. Freezeout temperatures obtained from the STAR~\cite{Adamczyk:2017iwn} and ALICE~\cite{Andronic:2017pug} collaborations are also shown.}
	\end{center}
	\label{fig:Tpc}
\end{figure}

When approaching the critical end point from the small $\mu_B$ side the transition strength should become stronger. As seen from the left plot of Fig.~\ref{fig:Tpc} the disconnected chiral susceptibility barely changes at $\mu_B$ up to around 250 MeV~\cite{Bazavov:2018mes}. Similar observation from chiral condensates using the imaginary chemical approach from the Wuppertal-Budapest collaboration shows that the transition strength does not increase up to $\mu_B$ around 300 MeV~\cite{Borsanyi:2020fev}. Thus combing with previous results, and recent lattice studies on the radius of convergence~\cite{Bazavov:2017dus,DElia:2016jqh}, the best estimate on the location of CEP from  lattice QCD computations is  
\begin{eqnarray}
T_c^{CEP} < 135-140~ \mathrm{MeV},~~~~ \mu_B^{CEP} > 300 ~\mathrm{MeV}.  
\end{eqnarray}

At vanishing baryon chemical potential $T_{pc}$ is determined as 156.5$\pm$1.5 MeV in the continuum limit based on the O(4) scaling analyses of various chiral observables~\cite{Bazavov:2018mes}, while it is 158.0$\pm$0.6 MeV as recently determined from the relation between the light quark condensate and its susceptibility~\cite{Borsanyi:2020fev}. At nonzero baryon density the chiral crossover transition temperature, which is parameterized as
\begin{equation}
T_{pc}(\mu_B) = T_{pc}(\mu_B=0) \left(1-\kappa_2\left(\frac{\mu_B}{T}\right)^2 - \kappa_4\left(\frac{\mu_B}{T}\right)^2\right),
\end{equation}
has been determined using the Taylor Expansion method~\cite{Bazavov:2018mes}  as well as the imaginary chemical potential approach~\cite{Borsanyi:2020fev,Bonati:2018nut,Bonati:2015bha,Bellwied:2015rza}.
The value of $\kappa_2$ is determined within the range of (0.008,0.0171) ~\cite{Bazavov:2018mes,Borsanyi:2020fev,Bonati:2018nut,Bonati:2015bha,Bellwied:2015rza}, and $\kappa_4$ is consistent with zero ~\cite{Bazavov:2018mes,Borsanyi:2020fev}.
The chiral crossover transition line at nonzero baryon chemical potential up to the order of $(\mu_B/T)^4$  obtained from  two most recent studies~\cite{Bazavov:2018mes,Borsanyi:2020fev} is shown in the right plot of Fig.~\ref{fig:Tpc}.

While to have the feasibility to probe the phase structure of QCD it would be convenient that the experimental freezeout line is close to the chiral crossover line. As shown in the right plot of Fig.~\ref{fig:Tpc}  the freezeout temperature at the LHC energy obtained from particle yields is in very good agreement with chiral crossover temperature at vanishing baryon chemical potential obtained from lattice QCD computations. On the other hand the freezeout temperatures at RHIC energies are also in good agreement to QCD transition line at $\mu_B$ larger than about 80 MeV, while the 200 GeV data point is about 1.5 sigma away from the transition line. 

\section{Skewness, kurtosis, hyper-skewness and hyper-kurtosis ratios of baryon number fluctuations}

Fluctuations of and correlations among conserved charges, like baryon numbers, electrical charge and strangeness have been long considered as sensitive observables to explore the QCD phase structure. Experimental proxies, like proton, kaons are used for measurements of mean, variance, skewness and kurtosis of conversed charges.  Theoretically in Lattice QCD, these are general susceptibilities, which are derivatives of pressure with respect to chemical potentials.  Expressions for skewness, kurtosis, hyper-skewness and hyper-kurtosis ratios of baryon number fluctuations are listed as follows
\begin{eqnarray}
\frac{S_B\,\sigma_B^3}{M_B}&=& \frac{\left\langle (\delta N_B)^3\right\rangle}{\left\langle N_B\right\rangle} = \frac{\chi_3^B(T,\mu_B)}{\chi_1^B(T,\mu_B)}=R_{31}^B(T,\mu_B),~\kappa_B\,\sigma_B^2=
\frac{\left\langle (\delta N_B)^4\right\rangle}{\left\langle (\delta N_B)^2\right\rangle} = \frac{\chi_4^B(T,\mu_B)}{\chi_2^B(T,\mu_B)}=R_{42}^B(T,\mu_B),\\
\frac{S^h_B\sigma_B^5}{M_B}\,=&=&\frac{\left\langle (\delta N_B)^5\right\rangle}{\left\langle N_B\right\rangle} = \frac{\chi_5^B(T,\mu_B)}{\chi_1^B(T,\mu_B)} =R_{51}^B(T,\mu_B),~~\kappa^h_B\sigma_B^4=\frac{\left\langle (\delta N_B)^6\right\rangle}{\left\langle (\delta N_B)^2\right\rangle} = \frac{\chi_6^B(T,\mu_B)}{\chi_2^B(T,\mu_B)} = R_{62}^B(T,\mu_B),
\end{eqnarray}
where $\chi_n^B(T,\mu_B)$ is the general susceptibility defined as the n-th order derivative of pressure with respect to baryon chemical potential $\mu_B$. On the lattice $\chi_n^B(T,\mu_B)$  can be computed by Taylor expanding the pressure obtained either at zero baryon chemical potential or at nonzero imaginary chemical potential up to a certain order of $\mu_B$. This renders the comparison of QCD results with experiment measurement possible~\footnote{Establishing the connection from the critical behavior of fluctuations of conserved charges in the chiral limit to that at the physical pion mass at vanishing baryon chemical potential in lattice QCD studies is important and also relevant for studies at LHC energies~\cite{Arslandok:2020mda}, and it has been pursed very recently~\cite{Sarkar:2019jwa}.}. Although many effects,  e.g. non-equilibrium effects, detector effects etc (see a recent review~\cite{Bzdak:2019pkr}), could spoil the comparison, a baseline for the thermal equilibrated QCD provided by Lattice QCD computations is certainly essential.

\begin{figure}[htp!]
	\centering
	\includegraphics[width=0.42\textwidth]{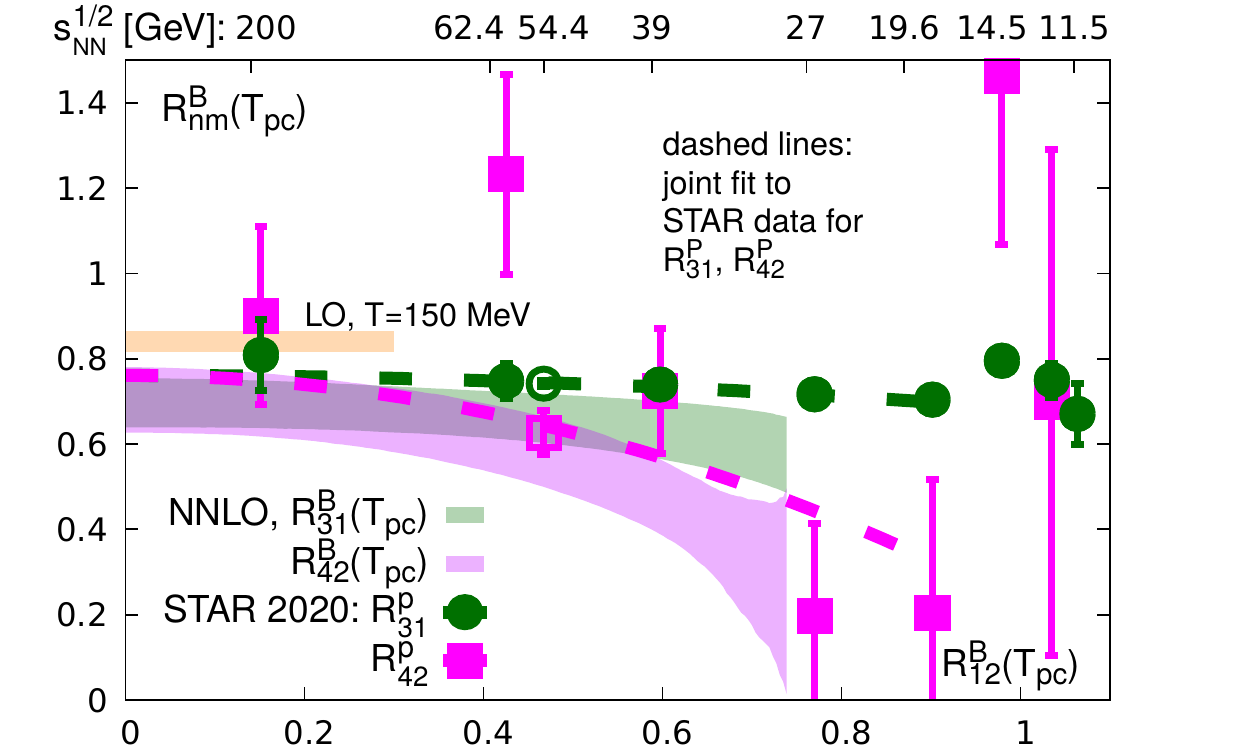}
	\includegraphics[width=0.42\textwidth]{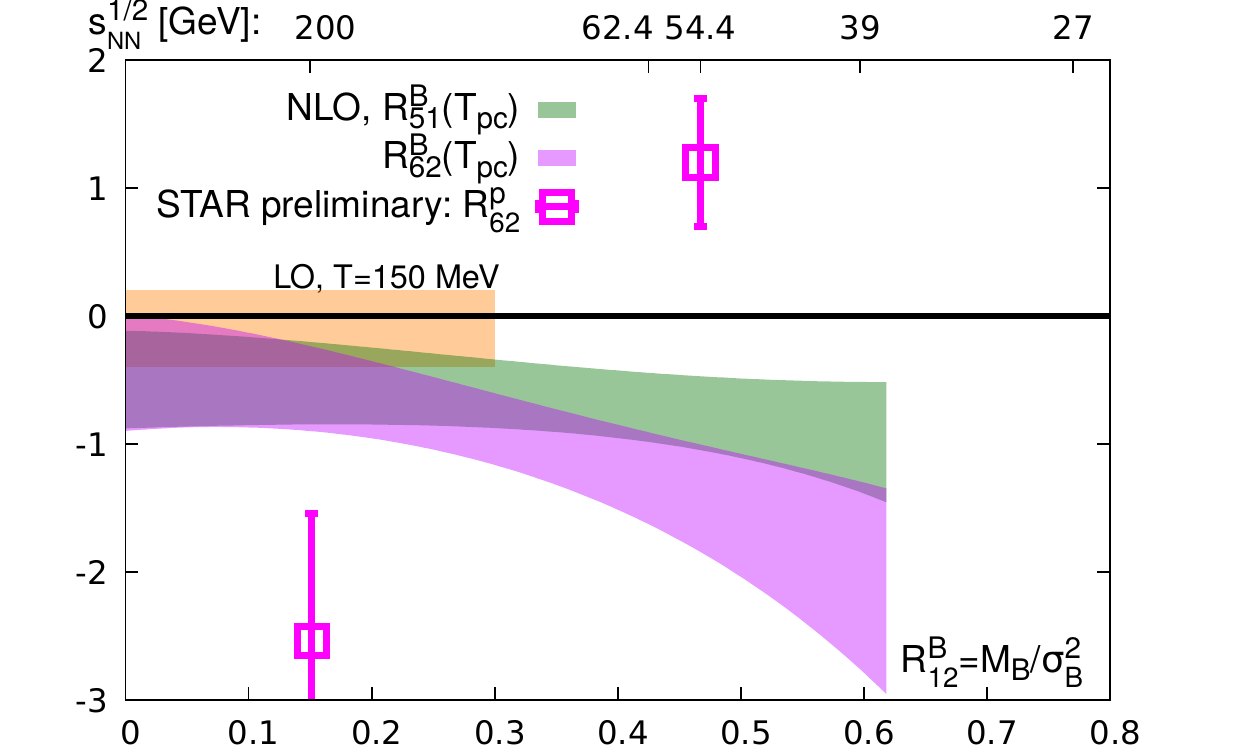}
	\caption{ Skewness ($R_{31}^B$) and
		kurtosis ($R_{42}^B$) ratios based on $N_t=8$ and 12 lattices computed up to NNLO Taylor series (left), and  hyper-kurtosis ($R_{62}^B$) and hyper-skewness ($R_{51}^B$) ratios of net baryon number fluctuations computed up to NLO Taylor series (right) as functions of $R_{12}^B(T,\mu_B) \equiv M_B/\sigma_B^2$ on the chiral crossover line obtained in lattice QCD ~\cite{Bazavov:2020bjn,Bollweg:2020yum}.
		The yellow bands denote the results for $R_{31}^B\approx R_{42}^B=0.84(2)$ (left), and $R_{51}^B\approx R_{62}^B=-0.1(3)$ (right) both at $\mu_B=0$ and T=150 MeV up to LO order Taylor series.
		 Square and circle data points are results on cumulant ratios of net proton-number
		 fluctuations obtained by the STAR Collaboration \cite{Adam:2020unf}.
		 Also shown are preliminary results obtained at
		 $\sqrt{s_{_{NN}}}=54.4$~GeV \cite{STARcumulants}.
		The STAR data points shown in the left plot are obtained in the centrality class of 0-5\% while those in the right plot in the centrality class of 0-40\%. Details can be found in Refs.~\cite{Bazavov:2020bjn,Bollweg:2020yum}.
			}
	\label{fig:cumulants}
\end{figure}

The skewness and kurtosis ratios of net baryon number fluctuations have been computed from LQCD up to next-next-to-leading order (NNLO) in $\mu_B$ where the computation of the 8th order Taylor expansion coefficients at vanishing $\mu_B$  is needed~\cite{Bazavov:2017tot,Bazavov:2020bjn}.
They are shown as a function of $R_{12}^B=\chi_1^B/\chi_2^B$, i.e. ratio of mean to variance at temperatures along the chiral crossover transition line $T_{pc}(\mu_B)$  in the left plot of Fig.~\ref{fig:cumulants}.  The green band showing the skewness ratio is almost flat, while the purple band denoting the kurtosis ratio $R_{42}(T_{pc})$ decreases faster than the skewness ratio. The trend for both skewness and kurtosis ratios obtained from the STAR collaboration is quite consistent with LQCD results.  There the preliminary results for both kurtosis and skewness ratios at $\sqrt{s_{NN}}=54.4$ GeV with high statistics are in good agreement with the LQCD computations~\cite{STARcumulants}, while those at $\sqrt{s_{NN}}=200$ GeV agree with LQCD results within errors~\cite{Adam:2020unf}. 
We also note that the STAR data at 62.4 GeV show an ordering
of the skewness and kurtosis data, which is opposite to what one
would expect from the LQCD results, and also is opposite to what
is found at 54.4 GeV. As the latter has much higher statistics, the
results at 62.4 GeV may suffer from too low statistics.
It is worth noting that at vanishing baryon chemical potential, $R_{31}\approx R_{42}$ is about 0.84 at a temperature e.g. 150 MeV (shown as the yellow band in the left plot of Fig.~\ref{fig:cumulants}), which is larger than their values at $T_{pc}$ and has better agreement with the STAR data point. Based on this observation one may suspect that the freezeout temperature $T_f$ at $\sqrt{s_{NN}}=$ 200  GeV could be lower than the transition temperature $T_{pc}$, however, this is in conflict with the freezeout temperature, $T_f\simeq164$ MeV, determined from
particle yields (see right plot of Fig.~\ref{fig:Tpc}). Both findings are thus thermodynamically not consistent.
Given the fact that at 54.4 GeV $T_f$ is in good agreement with the transition temperature as indicated from the right plot of Fig.~\ref{fig:Tpc} and the left plot of Fig.~\ref{fig:cumulants}, and the NNNLO contribution in lattice computations should be smaller at 200 GeV than at 54.4 GeV,  it is hard to understand the conflict in the freezeout temperature as obtained from the yields and kurtosis (skewness) ratios at 200 GeV. More investigations are needed. 

A first estimate of hyper-kurtosis and hyper-skewness ratios of net baryon number fluctuation from LQCD computations in the NLO order of $\mu_B$ has been presented in this year's QM conference~\cite{Bollweg:2020yum,Bazavov:2020bjn}. These two quantities are very much statistics hungry in both LQCD computations and experiment measurements. In LQCD computations the 8th order Taylor expansion coefficients are needed for the computation up to NLO and they in general have a very poor signal to noise ratio (cf. Fig. 1 in Ref.~\cite{Bazavov:2020bjn}), while the preliminary STAR results shown in the right plot of Fig.~\ref{fig:cumulants} have statistics of the order of 200 Million and 500 Million events for $\sqrt{s_{NN}}=200$ GeV and 54.4 GeV, respectively, and the results are obtained within the centrality class of 0-40\%. As clearly seen in the right plot of Fig.~\ref{fig:cumulants} that both hyper-kurtosis (purple band) and hyper-skewness (green band) from LQCD computations are negative at $R_{12}\lesssim 0.6$, corresponding to systems having colliding energies $\sqrt{s_{NN}}\gtrsim 39$ GeV. Although a nice consistency between QCD results and experimental data on skewness and kurtosis  ratios at 54.4 GeV is observed, 
large differences are seen in the determination of the hyper-kurtosis, i.e.  the sign of hyper-kurtosis obtained from LQCD and the STAR experiment is different. To turn the QCD results to be positive and consistent with the current experiment data  a sufficiently large NNLO contribution (10th order Taylor expansion coefficients) is needed. However, as commonly expected that at $\sqrt{s_{NN}}=$54.4 GeV the system is not in the proximity to the critical end point a large NNLO contribution may not occur.
In the case of  $\sqrt{s_{NN}}=200$ GeV a similarly large difference is also observed between LQCD results and the experiment data although the signs are the same as seen in the right plot of Fig.~\ref{fig:cumulants} . Since $\mu_B$ is smaller at $\sqrt{s_{NN}}$=200 GeV than that at 54.4 GeV the NNLO contributions should be less significant at $\sqrt{s_{NN}}$=200 GeV. Thus to better understand the discrepancy at $\sqrt{s_{NN}}$=200 GeV more statistics in the experiment is needed, while for the case of $\sqrt{s_{NN}}=54.4$ GeV a better understanding of NNLO contributions in the LQCD computations would be required to provide a more precise baseline for a equilibrated QCD medium. 

Since the signal to noise ratio is much better in the hyper-skewness ratio compared to that in the hyper-kurtosis ratio, it would be interesting to see the results of hyper-skewness ratios from the STAR collaboration. Similar as lower cumulant ratios, i.e. skewness and kurtosis, hyper-skewness and hyper-kurtosis ratios are roughly the same at vanishing $\mu_B$, and they are shifted upwards at lower temperatures, e.g.  $R_{51}\simeq R_{62}=-0.1\pm0.3$ at T=150~MeV including contributions up to LO Taylor series shown as the yellow band in the right plot of Fig.~\ref{fig:cumulants}. 
\vspace{-0.2cm}
\section{Quarkonia,  and charm and bottom quark drag coefficients}
\vspace{-0.2cm}
Quarkonia are considered as a thermometer of medium produced in heavy-ion collisions and the fate of quarkonia states in the medium as well as transport coefficients are among the main topics in LQCD studies in extreme conditions~\cite{Ding:2015ona,Rothkopf:2019ipj}. Many new theoretical results obtained since last years' QM conference is overviewed in Ref.~\cite{Rothkopf:2020vfz}. Following two interesting studies which were presented in this year's QM conference are worthy to be highlighted. 

\begin{figure}[hbtp!]
	\centering
	\includegraphics[width=0.33\textwidth]{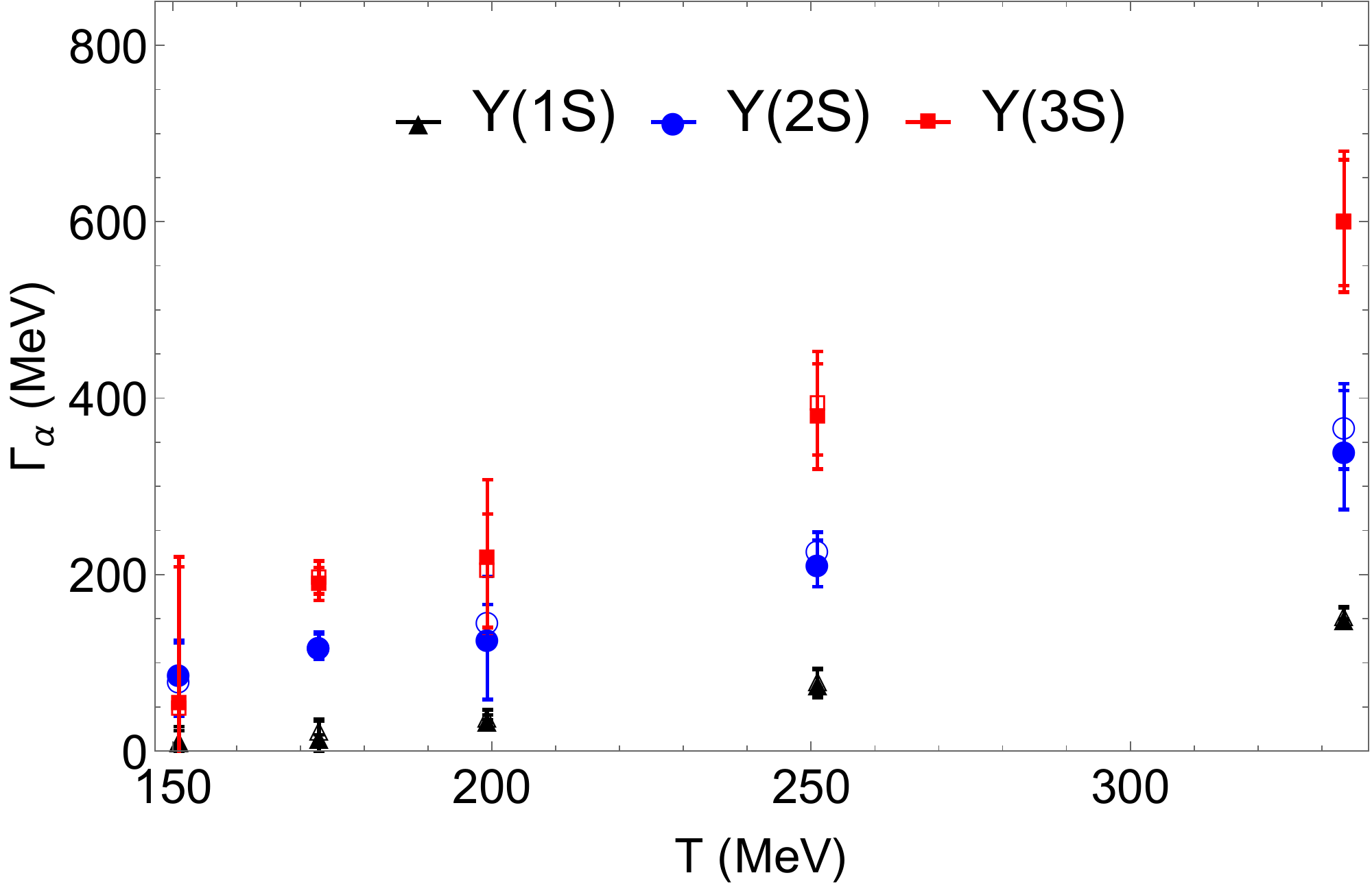}~\includegraphics[width=0.33\textwidth]{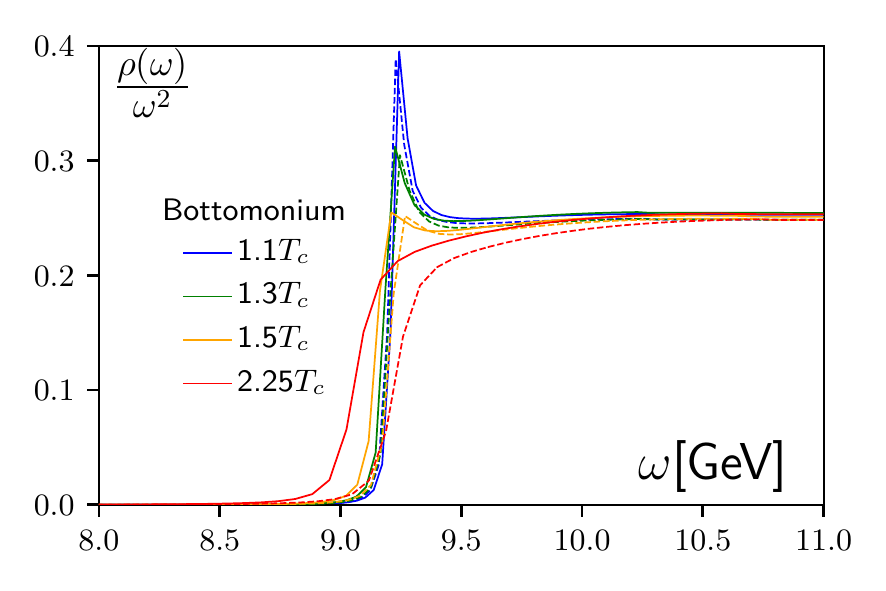}
	~\includegraphics[width=0.28\textwidth]{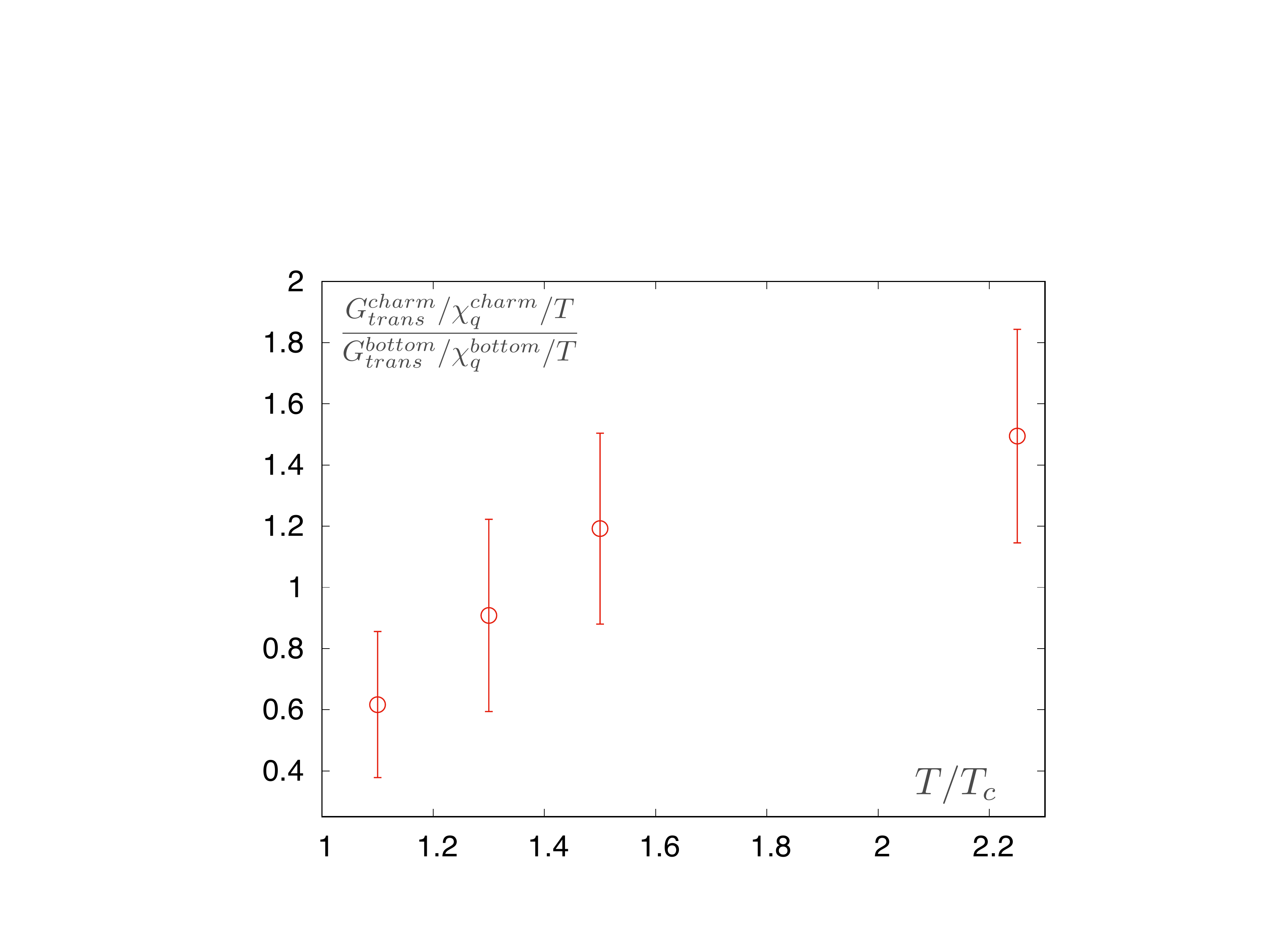}
	\caption{Left: Thermal width of $\Upsilon$ as obtained using two models (open and filled points) to fit the correlation functions computed in the framework of NRQCD on $48^3\times 12$ lattices with $m_\pi\approx 161$ MeV. Details can be found in Ref.~\cite{Larsen:2019zqv}. Middle: Spectral functions of bottomonia in the vector channel in quenched QCD at various temperatures above the transition temperature from fits to continuum-extrapolated correlators using a pNRQCD motivated ansatz~\cite{Lorenz:2020uik}. Solid lines denote fit results while dotted lines represent the original pNRQCD motivated ansatz. Right: Ratio of transport contributions between those from charm ($G_{trans}^{charm}$) and bottom ($G_{trans}^{bottom}$) quarks normalized by their corresponding quark number susceptibility ($\chi_q^{charm,bottom}$) and temperature. See more in the text and in Ref.~\cite{Lorenz:2020uik}.}
	\label{fig:HQ}
\end{figure}

A first lattice study on excited bottomonia  states up to 3S and 2P state at nonzero temperature has been carried out in the framework of NRQCD on $48^3\times 12$ lattices with $m_\pi\approx 161$ MeV~\cite{Larsen:2019zqv}. The excited bottomonium states at nonzero temperatures were probed through temporal correlation functions using  extended operators which have good overlaps with excited vacuum bottomonia~\cite{Larsen:2019bwy}. The thermal width of bottmonium states are obtained as fit parameters to correlation functions with two model spectral functions. As shown in the left plot of Fig.~\ref{fig:HQ} the open points denote thermal width of $\Upsilon$ states obtained using a model spectral function consisting of only $\delta$ functions while the filled points are from a model spectral function consisting also a Gaussian form. The obtained thermal width follows the pattern of
$\Gamma_{\Upsilon(1S)}(T) < \Gamma_{\chi_{b0}(1P)}(T) < \Gamma_{\Upsilon(2S)}(T)<\Gamma_{\chi_{b0}(2P)}(T) <\Gamma_{\Upsilon(3S)}(T)$,
which is compatible with the sequential dissociation picture of quarkonia.

A first continuum extrapolated results of $J/\psi$ and $\Upsilon$ correlation functions have been obtained in quenched lattice QCD very recently~\cite{Lorenz:2020uik}. The continuum extrapolation makes the direct comparison with pQCD computations possible, and thus  alternatively from using inversion methods~\cite{Rothkopf:2019ipj,Asakawa:2020hjs,Rothkopf:2020qqt}, the spectral functions are investigated by fitting to the correlators using a pQCD motivated ansatz~\cite{Burnier:2017bod}. It was found that no $J/\psi$ peak is needed in the fit ansatz to describe the lattice correlator data $T\geq1.1 T_c$, while a resonance peak is still needed for $\Upsilon$ up to 1.5 $T_c$ as shown in the middle plot of Fig.~\ref{fig:HQ}.

The right plot of Fig.~\ref{fig:HQ} shows the ratio of transport contributions in the bottomonium correlation functions to that in the charmonium correlation functions. These transport contributions are obtained by subtracting the correlators resulted from fit spectral functions (solid lines in the middle plot of Fig.~\ref{fig:HQ}) from the lattice correlator data at the middle point $\tau T=1/2$. Assuming a Lorentz ansatz for the transport peak the ratio of the transport contribution from charm quark to that from bottom quark normalized by their corresponding quark numbers susceptibility  $\frac{G^{charm}_{trans}/\chi^{charm}_q/T}{G^{bottom}_{trans}/\chi^{bottom}_q/T}\approx\frac{M_{b}}{M_{c}}~\frac{\mathrm{tan}^{-1}(T/\eta_{c})}{\mathrm{tan}^{-1}(T/\eta_{b})}$.
Here $\eta_c$($\eta_b$) denotes the drag coefficient for charm (bottom) quark. Since the mass of bottom quark $M_b$ is about 3 times of charm quark mass $M_c$, then $\mathrm{tan}^{-1}(T/\eta_{c})/\mathrm{tan}^{-1}(T/\eta_{b})$ turns out to be smaller than 1 as inferred from the right plot of Fig.~\ref{fig:HQ}. This suggests a flavor hierarchy in drag coefficients having $\eta_c > \eta_b$. 

\vspace{-0.5cm}
\section{Summary}
\vspace{-0.3cm}
Since the last Quark Matter conference considerable progress has been made in understanding the equilibrium thermodynamics of strong interaction matter through lattice QCD calculations.
Some highlights are:  1) The state-of-the-art estimate on the location of CEP is $T_c^{CEP} < 135-140~ \mathrm{MeV}$ and $\mu_B^{CEP} > 300 ~\mathrm{MeV}$; 2) The kurtosis and skewness ratios at $\sqrt{s_{NN}}=54.4$ GeV measured from the STAR collaboration are in good agreement with QCD in thermal equilibrium while large differences are found for hyper-kurtosis ratios. It would be interesting to make the comparison for hyper-skewness ratios; 3) a sequential dissociation pattern of quarkonia states is found in the framework of lattice NRQCD, and a flavor hierarchy in quark drag coefficients is suggested from studies in quenched QCD.

\noindent\textbf{Acknowledgements}~ 
The work is supported by the National Natural Science Foundation of China under grant numbers 11535012 and 11775096.
\vspace{-0.5cm}
\bibliographystyle{elsarticle-num}
\vspace{-0.1cm}
\bibliography{HTD_refs}







\end{document}